\documentclass[12pt]{iopart}

\usepackage[utf8]{inputenc}

\usepackage[colorlinks,linkcolor=blue,citecolor=blue]{hyperref}

\usepackage{aas_macros}
\usepackage{graphicx}
\usepackage{float}
\makeatletter
\newcommand*{\rom}[1]{\expandafter\@slowromancap\romannumeral #1@}
\newcommand{\comm}[1]{}

\newcommand{\orcid}[1]{\href{https://orcid.org/#1}{\,\includegraphics[width=8px]{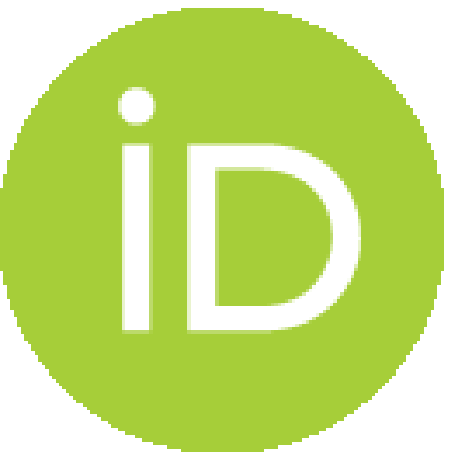}}}

\begin{document}

\title[21 cm Power Spectrum]{21 cm Power Spectrum for Bimetric Gravity and its Detectability with SKA1-Mid Telescope}

\author{Ajay Bassi$^{1,a}$\orcid{0000-0001-8915-3860}, Bikash R. Dinda$^{2,b}$ \orcid{0000-0001-5432-667X} \& Anjan A. Sen$^{3,a}$\orcid{0000-0001-9615-4909}}

\address{$^a$Centre for Theoretical Physics,  Jamia Millia Islamia, New Delhi (India) - 110025 }
\address{$^b$Department of Physical Sciences, Indian Institute of Science Education and Research Kolkata, Mohanpur, Nadia, West Bengal 741246, India}
\ead{$^1$\mailto{ajay@ctp-jamia.res.in},$^2$\mailto{bikashdinda.pdf@iiserkol.ac.in}, $^3$\mailto{aasen@jmi.ac.in}}
\vspace{10pt}
\begin{indented}
\item[]November 2022
\end{indented}

\begin{abstract}
We consider a modified gravity theory through a special kind of ghost-free bimetric gravity, where one massive spin-2 field interacts with a massless spin-2 field. In this bimetric gravity, the late time cosmic acceleration is achievable. Alongside the background expansion of the Universe, we also study the first-order cosmological perturbations and probe the signature of the bimetric gravity on large cosmological scales. One possible probe is to study the observational signatures of the bimetric gravity through the 21 cm power spectrum. We consider upcoming SKA1-mid antenna telescope specifications to show the prospects of the detectability of the ghost-free bimetric gravity through the 21 cm power spectrum. Depending on the values of the model parameter, there is a possibility to distinguish the ghost-free bimetric gravity from the standard $\Lambda$CDM model with the upcoming SKA1-mid telescope specifications.
\end{abstract}

\section{Introduction}
\label{intro}
Supernovae type Ia observations in 1998 \cite{SupernovaSearchTeam:1998fmf,SupernovaCosmologyProject:1998vns} revealed that the present expansion of the universe is in an accelerating phase. It had a very profound effect on our understanding of the universe as no ordinary or dark constituents (with an attractive gravitational force) can account for this accelerated expansion, with the Einstein's general theory of relativity (GR) as the standard theory of gravitation. Even two decades later, we don't have any firmly established theoretical explanation for this phenomenon. Two approaches have been adopted in the past to address this problem. The first approach is to introduce an exotic fluid with negative pressure dubbed as 'dark energy' in the energy budget of the universe \cite{Linder:2008pp,Sahni:1999gb,Silvestri:2009hh}, which can produce the desired repulsive gravitational effect to source the accelerated expansion of the universe and the second approach is to modify the standard theory of gravitation at large cosmological scales. 

The cosmological constant $\Lambda$ is the simplest possible candidate for the dark energy \cite{Peebles:2002gy,Padmanabhan:2002ji,Carroll:2000fy}. The cosmological constant $\Lambda$ along with cold dark matter 'CDM', also known as the concordance $\Lambda$CDM model, has been very successful to explain most of the cosmological observations \cite{Planck:2018vyg}, but it is plagued with certain theoretical issues like fine tuning problem \cite{Martin:2012bt} and cosmic coincidence problem \cite{Zlatev:1998tr}. Alongside these well known theoretical issues, some cosmological observations, such as the model independent local measurement of the Hubble constant $H_0$ \cite{Riess:2019cxk} and the direct measurement of the fluctuations of matter density distribution in the universe ($S_8$) \cite{Asgari:2019fkq}, are in tension with the measurement of these parameters by Planck observations of cosmic microwave background (CMB) with $\Lambda$CDM as the underlying model \cite{Planck:2018vyg}. To counter these issues, time evolving dark energy models have been studied in the past and we refer \cite{Copeland:2006wr} for a comprehensive review of these models. 

Apart from dark energy models, modified theories of gravity \cite{Tsujikawa:2010zza} have been used to explain the late time cosmic acceleration without taking into account any unknown form of dark energy or dark matter. One has to be careful that the modified theory in consideration should restore the general relativity on small scales because GR's predictions of gravitational observations on solar system scales are astonishingly precise \cite{Will:2014kxa}. The first attempt to modify GR by introducing mass to the intermediate particle for the gravitational force, the graviton, through linear theory of massive gravity was done by Fierz and Pauli \cite{Fierz:1939ix}. But this theory contains Boulware-Desert (BD) Ghost \cite{Boulware:1972yco}. This BD ghost can be removed by inclusion of a second metric into the theory alongside the physical metric $g_{\mu\nu}$ with carefully constructed interaction term between these two metrics \cite{deRham:2010kj}. Dynamics of the second metric give rise to the bimetric gravity \cite{Hassan:2011zd}. The non-standard background cosmology of bimetric gravity can lead to the late time cosmic acceleration without any explicit dependence on dark energy \cite{Volkov:2011an,vonStrauss:2011mq}. The bimetric gravity has a screening mechanism that can restore the general relativity on solar system scales \cite{Enander:2015kda,Babichev:2013pfa}.

A good modified gravity model or dark energy model should be consistent with the different cosmological observations. The observation of 21 cm emission of the neutral HI gas can be a good tracer for the underlying dark matter distribution. After the completion of reinonization epoch at redshift $z\sim 6$, the universe was mostly ionized \cite{Fan:2001vx,SDSS:2001tew}. The bulk of neutral hydrogen HI gas is thought to be residing inside the self-shielded damped Ly$\alpha$ (DLA) systems \cite{Wolfe:2005jd}, where it is shielded from the ionizing UV photons. The detection of individual DLA clouds can be a mammoth task, due to their weak signal, but fortunately this is not necessary as one can measure the collective diffused HI intensity over all the DLA clouds at large scales. This forms a background radiation in low-frequency radio observations (frequencies $<$ 1420 MHz), similar to the background CMBR, except that the signal here is a function of redshift because observations at different frequencies probe the HI intensity at different distances. This background HI radiation is a biased tracer for the matter distribution and the HI power spectrum is related to matter power spectrum through HI brightness temperature. 

The HI power spectrum on large scales can be a powerful tool to study the large scale structure formation \cite{Loeb:2008hg,Wyithe:2007gz}. The upcoming SKA1-Mid telescope is specifically designed to constrain the possible deviation from the general relativity on cosmological scales by measuring the large scale distribution of neutral hydrogen HI over the low redshift regions (z $\sim$ 3), which can map the structure of the universe on large scales \cite{Braun:2019gdo,SKA:2018ckk}. This mapping will help us to distinguish between different modified gravity models and dark energy models. The aim of this paper is to study the 21 cm power spectrum for the bimetric gravity and check the detectability of deviation in this model from the $\Lambda$CDM model in the context of the forthcoming SKA1-Mid telescope specifications.

\section{Bimetric gravity}
\label{bim-grav}

In ghost-free bimetric gravity, the acceleration of the universe is achieved by the interaction of two symmetric spin-2 fields (i.e. metrics): one is the physical metricc denoted by $g_{\mu\nu}$ and another one is auxiliary metric denoted by $f_{\mu\nu}$. In bimetric gravity, we assume that there is only one matter sector, represented by matter Lagrangian ${L}_m$, coupled to the physical metric $g_{\mu\nu}$. In this scenario, the action (called the Hassan-Rosen action \cite{Hassan:2011zd,Mortsell:2018mfj}) is given by

\begin{equation}
S = \int d^4x\left[\sqrt{-det~g} \left(\frac{R}{2\kappa_g} + \sum_{n=0}^4\beta_{n}e_n\left(\sqrt{g^{-1}f}\right) + L_m\right)+ \sqrt{-det~f} \frac{\tilde{R}}{2\kappa_f} \right],
\label{eq:action}
\end{equation}

\noindent
where $R$ and $\tilde{R}$ are the Ricci scalars corresponding to the metrics $g_{\mu\nu}$ and $f_{\mu\nu}$, respectively. $\kappa_g$ and $\kappa_f$ are the gravitational constants corresponding to the metrics $g_{\mu\nu}$ and $f_{\mu\nu}$, respectively. $\beta_n$'s are five constants and $e_n$s are the five elementary symmetric polynomials of the eigenvalues of the matrix $\sqrt{g^{-1}f}$ \cite{Hassan:2011zd} ($\forall \hspace{0.1 cm} n \in [0,1,2,3,4]$). $d^4x$ is the usual four-volume element.

The second term insider the first bracket in Eq.~\ref{eq:action} gives the interaction between two metrics whereas the last term inside the first bracket in Eq.~\ref{eq:action} gives the coupling between matter and metric $g_{\mu\nu}$.

The five $\beta$ parameters are considered to characterise different cosmological aspects for the bimetric gravity. These parameters are not unique because of possible re-scaling under which the Hassan-Rosen action in Eq.~\ref{eq:action} remains invariant \cite{Luben:2020xll}. We use dimensionless, re-scaling invariant parameters from M\"{o}rtsell et.al. \cite{Mortsell:2018mfj} given by

\begin{equation}
    B_i\equiv \frac{\kappa_g \beta_i}{H_0^2},
\end{equation}

\noindent
$\forall \hspace{0.1 cm} i \in [0,1,2,3,4]$, where $H_0$ is the present value of the Hubble parameter. Various models with different combinations of nonzero $B_i$'s have been studied in the past \cite{Mortsell:2018mfj,Hogas:2021fmr,Hogas:2021lns,Mortsell:2017fog}. We have used the model \cite{Dhawan:2020xmp} in which only $B_0$ and $B_1$ are nonzero, and are related by

\begin{equation}
    B_0 = 3\left( 1 - \Omega_m - \frac{B_1^2}{3}  \right),
    \label{eq:b0}
\end{equation}

\noindent
where $\Omega_m$ is the present value of the matter energy density parameter. The resulting modified Friedmann equation is given as

\begin{equation}
    \frac{H^2}{H_0^2} = \frac{\Omega_m}{2 a^3} +\frac{B_0}{6} +
            \sqrt{\left( \frac{\Omega_m}{2 a^3} +\frac{B_0}{6}  \right)^2 
                +\frac{B_1^2}{3}},
     \label{eq:fried}           
\end{equation}

\noindent
where $a$ is the scale factor.

\section{Large scale structure growth}

We have used the formalism by Mulamaki et al. \cite{Multamaki:2003vs} to study the large scale structures growth for the bimetric gravity. In this regard, we should mention that perturbations at linear scale in bimetric gravity gravity are suffered from gradient instability on sub horizon scales in the early time as well as Higuchi Ghost as the Hubble scale exceeds that effective gradient mass $m_{eff}$ \cite{Koennig_2014a,Koennig_2014b,Koennig_2015,Comelli_2014,Felice_2014,Akrami_2015}. This can be problematic for setting initial conditions to solve the evolution equation for the density contrast. There can be nonlinear effects  \cite{Aoki_2015} in the scalar graviton mass that can solve this issue. But these instability issue has not settled completely and needs further investigations. For detail discussion in this regard, we refer the interested reader to one recent work by bassi et al. \cite{bassi2023cosmological} We carry forward our calculations for solving the linear perturbation equations assuming that there are some nonlinear effects that can solve this issue of instability.

\noindent
Raychaudhuri's equation for a shear-free and irrotational fluid, with four-velocity $u^\mu$, is given by

\begin{equation}
    \dot{\Theta} + \frac{\Theta^2}{3} = R_{\mu\nu} u^\mu u^\nu,
    \label{eq:rayc}
\end{equation}

\noindent
where $\Theta = \nabla_\mu u^\mu$ and $R_{\mu\nu}$ is the Ricci tensor corresponding to the metric $g_{\mu\nu}$. $\nabla$ is the usual gradient operator and over-dot represents the derivative with respect to the physical time $t$. We choose a coordinate system in which the four-velovity $u^{\mu}$ is given as

\begin{equation}
    u^{\mu}=(1,\dot{a} \bf{x}+\bf{v}),
    \label{eq:4vel}
\end{equation}

\noindent
where $\bf{v}$ is the peculiar velocity three-vector and $\bf{x}$ is the spatial coordinate vector. This gives

\begin{equation}
    \Theta = 3\frac{\dot{a}}{a}+\frac{\theta}{a},
    \label{eq:theta}
\end{equation}

\noindent
where $\theta = \nabla.\bf{v}$. Using Eqs.~\ref{eq:4vel} and~\ref{eq:theta}, Eq.~\ref{eq:rayc} can be related to background Hubble expansion $\bar{H}$ and perturbed Hubble expansion $H$ (which is same notation used in Eq.~\ref{eq:fried} but from here on it is used to address the perturbed Hubble expansion) as

\begin{equation}
    \frac{\dot{\theta}}{a}+\frac{\theta}{a}\bar{H}+\frac{\theta^2}{3a^2}
 = 3(\dot{H}+H^2-\dot{\bar{H}}-{\bar{H}}^2).   
 \label{eq:thetaray}
 \end{equation}

 \noindent
Using the perturbed continuity equation for non-relativistic matter, Eq.~\ref{eq:thetaray} in a matter dominated universe can be written as

\begin{equation}
    \delta^{\prime\prime}+\left( 2 + \frac{\dot{\bar{H}}}{\bar{H}^2} \right)\delta^{\prime} - \frac{4}{3}\frac{1}{1+\delta}(\delta^{\prime})^2 = -3\frac{1+\delta}{\bar{H}^2}\left[\left( \dot{H} + H^2 \right) -\left( \dot{\bar{H}} + \bar{H}^2 \right)  \right],
    \label{eq:delray}
\end{equation}

\noindent
where $\delta$ is local matter density contrast and prime is derivative w.r.t. $\ln({a})$. Quantities with overhead bar are the background quantities. We can expand the $\dot{H}$+$H^2$ term in terms of $\delta$ as

\begin{equation}
    3\frac{1+\delta}{\bar{H}^2}\left[\left( \dot{H} + H^2 \right) -\left( \dot{\bar{H}} + \bar{H}^2 \right)  \right] 
    = 3(1+\delta)\sum_{n=1} c_n \delta^n
    \label{eq:hindel}
\end{equation}

\noindent
 where $c_n$ are the coefficients for the matter perturbations $\delta^n$ (order $n$). Using above equation in Eq.~\ref{eq:delray}, we can get perturbation equations at linear and higher orders. We expand $\delta$ as

\begin{equation}
    \delta = \sum_{i=1}^\infty \frac{D_i(\eta)}{i!}\delta_0^i
\end{equation}

\noindent
where $\delta_0^i$'s are the small perturbations and the expansion parameter and $D_i$'s are the growth functions. Using this in Eq.~\ref{eq:delray}, we get the linear order perturbation equation as

\begin{equation}
    D_1^{\prime\prime} + \left(2+\frac{\dot{\bar{H}}}{\bar{H}^2} \right)D_1^\prime
    +3c_1D_1 = 0,
    \label{eq:1pert}
\end{equation}

\noindent
where prime represents derivative w.r.t. $\ln(a)$. Using the Hubble expansion Eq.~\ref{eq:fried} for the bimetric gravity, we get $c_1$ as

\begin{equation}
c_1 = \frac{\frac{-2\Omega_m}{a^3}\left(\frac{\Omega_m}{2a^3}+\frac{B_0}{6}\right)\left(\left(\frac{\Omega_m}{2a^3}+\frac{B_0}{6}\right)^2+\frac{B_1^2}{3}\right)-\frac{2\Omega_m}{a^3}\left(\left(\frac{\Omega_m}{2a^3}+\frac{B_0}{6}\right)^2+\frac{B_1^2}{3}\right)^{3/2}-\frac{\Omega_m^2B_1^2}{a^6}}{8\left(\frac{\Omega_m}{2a^3}+\frac{B_0}{6}+\sqrt{\left(\frac{\Omega_m}{2a^3}+\frac{B_0}{6}\right)^2+\frac{B_1^2}{3}}\right)\left(\left(\frac{\Omega_m}{2a^3}+\frac{B_0}{6}\right)^2+\frac{B_1^2}{3}\right)^{3/2}} .
\label{eq:c1}
\end{equation}

We solve Eq.~\ref{eq:1pert} to get the linear growth of the structure in the universe at large scales. We set the initial conditions at the decoupling epoch ($z=1000$), when universe was matter dominated. We consider the fact that at matter dominated epoch, $D_1\sim a$. 

Now we define linear matter power spectrum, using the linear growth function $D_1$, as
\begin{equation}
P_{m}(k,z) = A k^{n_{s}} T(k)^{2}~\frac{D_1^2(z)}{D_1^2(z=0)},
\label{mps}
\end{equation}

\noindent
where $k$ is the amplitude of the wave-vector $\bf{k}$, $A$ is the normalization constant fixed by $\sigma_8$ normalization, $n_s$ is the spectral index for the primordial density fluctuations and $T(k)$ is the transfer function given by Eisenstein and Hu \cite{Eisenstein:1997ik}. We fix the values of $\Omega_{\rm m0}=0.3111$, $\Omega_{\rm b0}=0.049$, $h=0.6766$, $n_s=0.9665$ and $\sigma_8=0.8102$ according to the best fit values of the Planck 2018 results \cite{Planck:2018vyg}. Here $\Omega_m$ is the present value of the matter energy density parameter, $\Omega_{\rm b0}$ is the present value of baryonic matter energy density parameter and $h$ is defined by the present value of Hubble parameter as $H_0 = 100~h~{\rm Km}S^{-1}{\rm Mpc}^{-1}$.

\section{21 CM Power Spectrum}
\label{21-cm-ps}
We don't observe dark matter distribution in the universe directly. The 21 cm emission in the post re-ionization epoch can be a good tracer of underlying dark matter distribution. Here we study the observational validity of bimetric gravity over $\Lambda$CDM model by using the 2-point correlation in the fluctuation of the excess HI 21 cm brightness temperature. The mean excess HI 21 cm brightness temperature is given by \cite{Datta:2006vh,Sarkar:2011jf}

\begin{equation}
C_{T} (z) = b \bar{x}_{\rm HI} \bar{T} (z),
\label{eq:ct}
\end{equation}

\noindent
where $b$ is the linear bias which connects the HI distribution to the underlying dark matter distribution, $\bar{x}_{\rm HI}$ is the mean neutral hydrogen fraction and $\bar{T} (z)$ is given by \cite{Datta:2006vh,Sarkar:2011jf}

\begin{equation}
\bar{T} (z) = 4.0 {\rm mK} (1+z)^{2} \left( \frac{\Omega_{\rm b0} h^{2}}{0.02} \right) \left( \frac{0.7}{h} \right) \frac{H_{0}}{H(z)}.
\label{eq:tbar}
\end{equation}

The 21 cm power spectrum, $P_{\rm 21}$, for the excess brightness temperature is given by \cite{Datta:2006vh,Sarkar:2011jf,Hussain:2016fgg,Bharadwaj:2004it}

\begin{equation}
P_{\rm 21} (k,z,\mu) = C_{T}^{2} \left(1+\beta_{T} \mu^{2}\right)^{2} P_{m}(k,z),
\label{eq:p212D}
\end{equation}

\noindent
where $\mu =\hat{n}.\hat{k}= \cos{\theta}$, where $\theta$ is the angle between line of sight unit vector $\hat{n}$ and the unit wave vector $\hat{k}$. $\beta_T$ is defined as

\begin{equation}
\beta_T = \frac{f}{b},
\label{eq:betat}
\end{equation}

\noindent
where $f$ is the linear growth factor defined as $f=d(\ln D_1)/d(\ln a)$. We consider linear bias $b=1$ throughout the calculations. The $\mu$ averaged 21 cm power spectrum, $P_{\rm 21}$ is given by

\begin{equation}
P_{\rm 21}(k,z) = \int_{0}^{1} d\mu \hspace{0.2 cm} P_{\rm 21}(k,z,\mu),
\label{eq:p21avg}
\end{equation}

\noindent
where we keep the same notation, $P_{\rm 21}$ for the $\mu$ averaged 21 cm power spectrum.

\begin{figure}[H]
  \centering
  \includegraphics[width=0.45\textwidth]{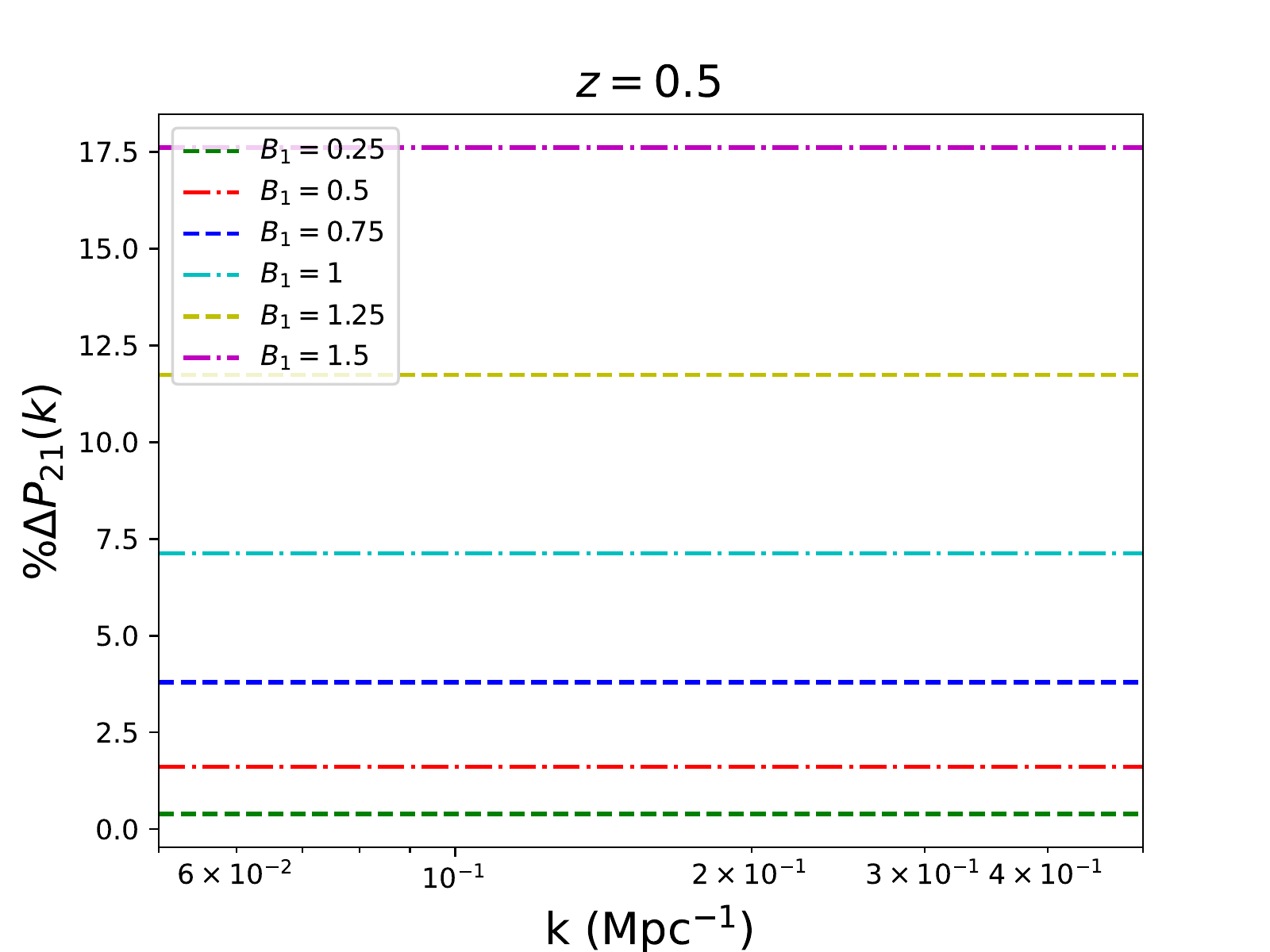}
  \includegraphics[width=0.45\textwidth]{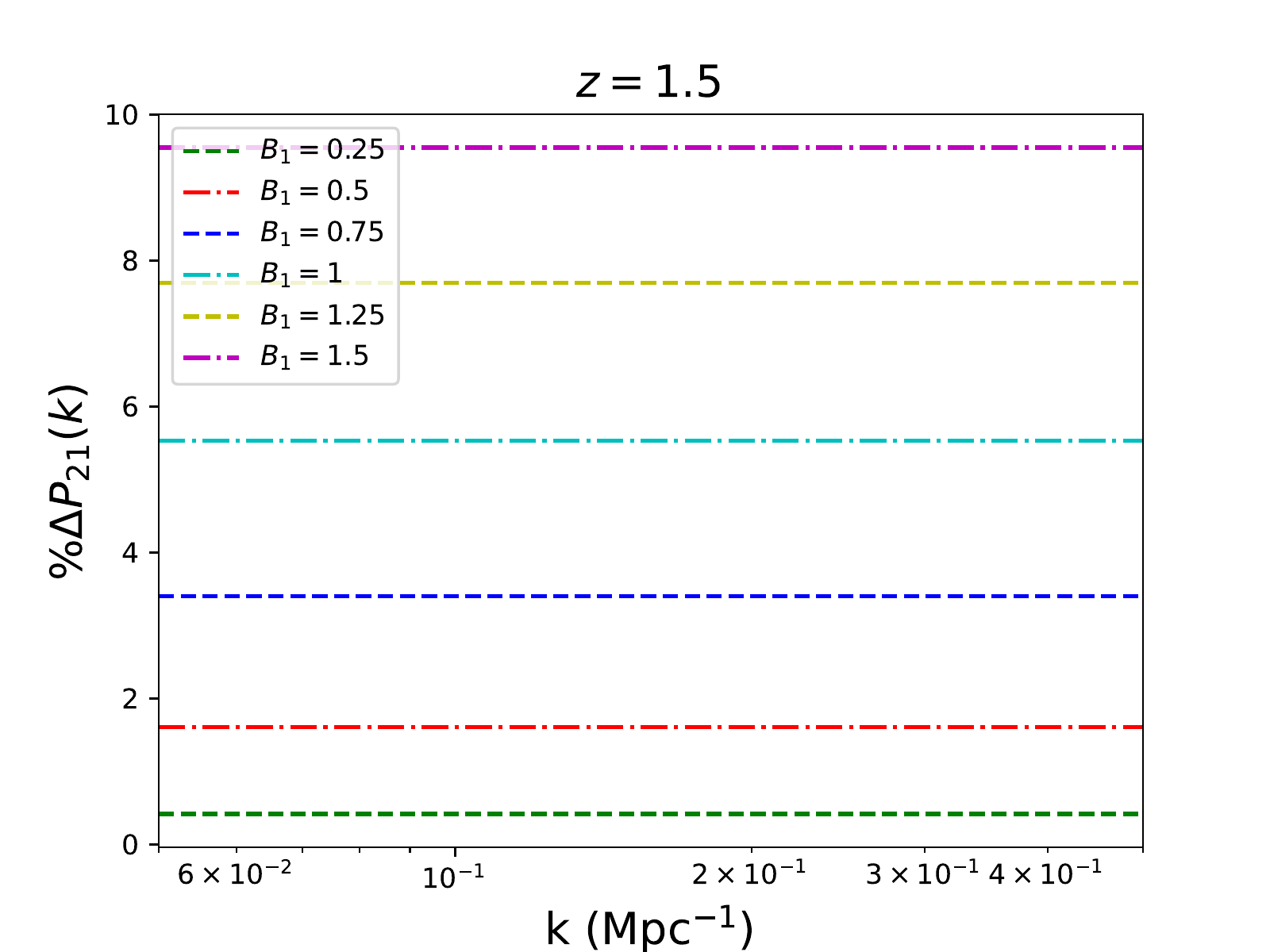}
  \includegraphics[width=0.45\textwidth]{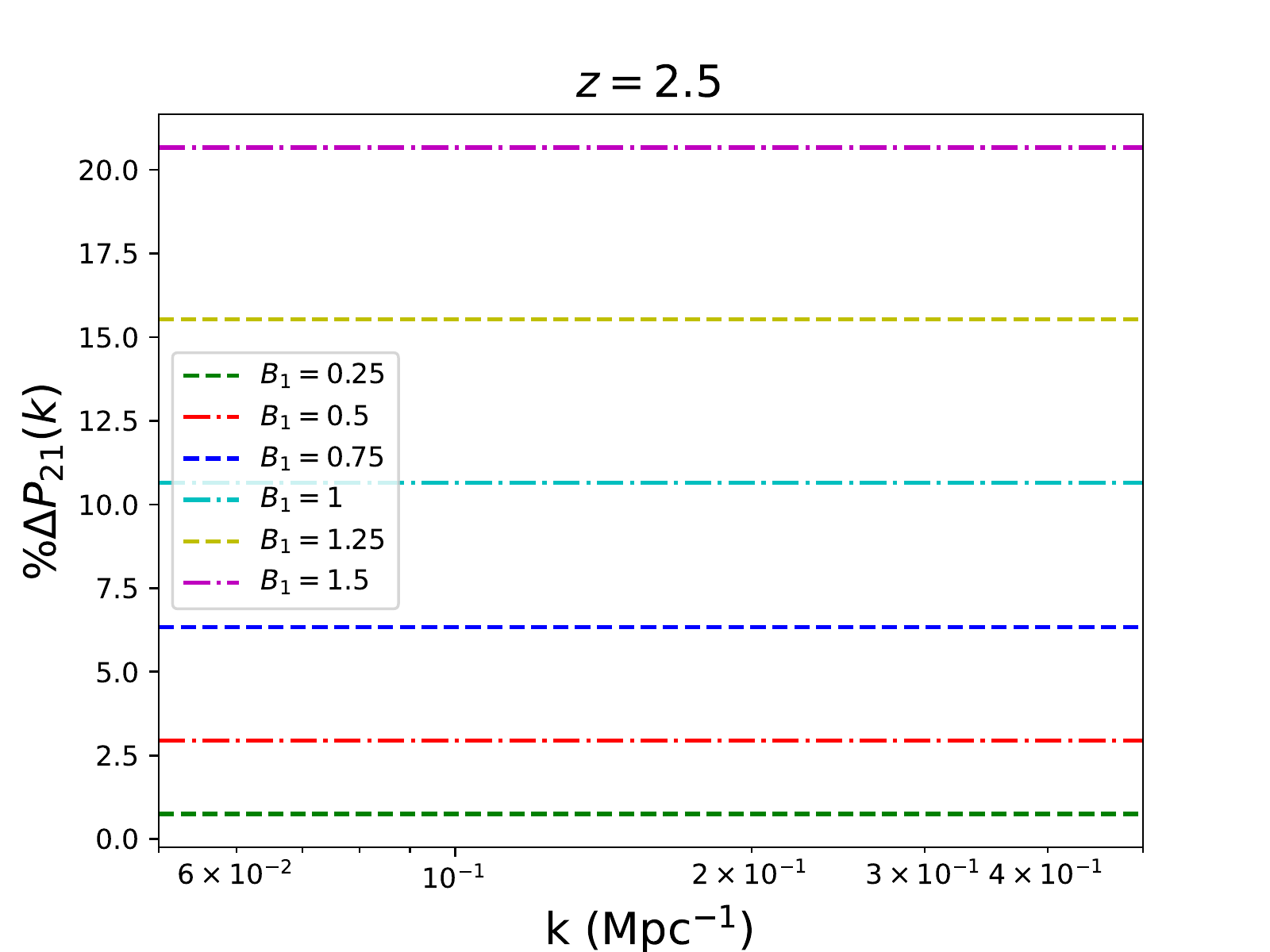}
  \caption{Percentage deviation in the 21 cm power spectrum for the bimetric gravity model from the $\Lambda$CDM model. We have plotted  $\%\Delta P_{\rm 21}$ w.r.t. $k$ at different redshifts for different values of parameter $B_1$. } 
  \label{fig:21_plot}
\end{figure}

In Fig.~\ref{fig:21_plot}, we show the absolute percentage deviation of the $\mu$ averaged 21 cm power spectrum from the $\Lambda$CDM model defined as $\%\Delta P_{\rm 21} = \left|P_{\rm 21}/P_{\rm 21}({\rm \Lambda CDM}) - 1\right| \times 100$. We see deviation of 0.5\% to 20\%, depending upon parameter $B_1$, at different redshifts.

21 cm HI emission falls well into the radio spectrum and is mostly immune to the obscuration by the intervening matter. The radio interferometers like SKA can measure the HI intensity over comparatively large angular scales \cite{Bull:2014rha}. The angular diameter distances and Hubble expansion rate as function of redshifts are measured by the baseline distributions of the interferometers using a fiducial model of cosmology. The difference between the real cosmological model and the fiducial cosmological model introduces additional anisotropies in the correlation function and we need to correct the 21 cm power spectrum given in Eq.~\ref{eq:p212D}. We have used the $\Lambda$CDM model as the fiducial model throughout the calculations. The observed 21 cm power spectrum for the bimetric gravity is given by \cite{Bull:2014rha,Ballinger:1996cd,Raccanelli:2015qqa}

\begin{eqnarray}
P_{\rm 21}^{\rm 3D}(k,z,\mu) &=& \frac{1}{\alpha_{||} \alpha_{\perp}^{2}} C_{T}^{2} \left[ 1+\beta_{T} \frac{\mu^{2} / F^{2}}{1+(F^{-2}-1) \mu^{2}} \right]^{2} \nonumber\\
&& \times P_{m} \left( \frac{k}{\alpha_{\perp}} \sqrt{1+(F^{-2}-1) \mu^{2}}, z \right).
\label{eq:p213D}
\end{eqnarray}

\noindent
Here we have used 3D superscript with 21 cm power spectrum notation because it is sometimes referred as 3D 21 cm power spectrum. In Eq.~\ref{eq:p213D}, $ \alpha_{||} = H_{\rm fd} / H $, $ \alpha_{\perp} = r / r_{\rm fd} $ and $ F = \alpha_{||} / \alpha_{\perp} $, where subscript 'fd' corresponds to the fiducial model and r is the line of sight comoving distance. The $\mu$ averaged 3D 21 cm power spectrum for bimetric gravity is given as

\begin{equation}
P_{\rm 21}^{\rm 3D}(k,z) = \int_{0}^{1} d\mu \hspace{0.2 cm} P_{\rm 21}^{\rm 3D}(k,z,\mu),
\label{eq:P213Davg}
\end{equation}

\noindent
where we keep the same notation, $P_{\rm 21}^{\rm 3D}$, for $\mu$ averaged 3D 21 cm power spectrum.

\begin{figure}[H]
  \centering
  \includegraphics[width=0.45\textwidth]{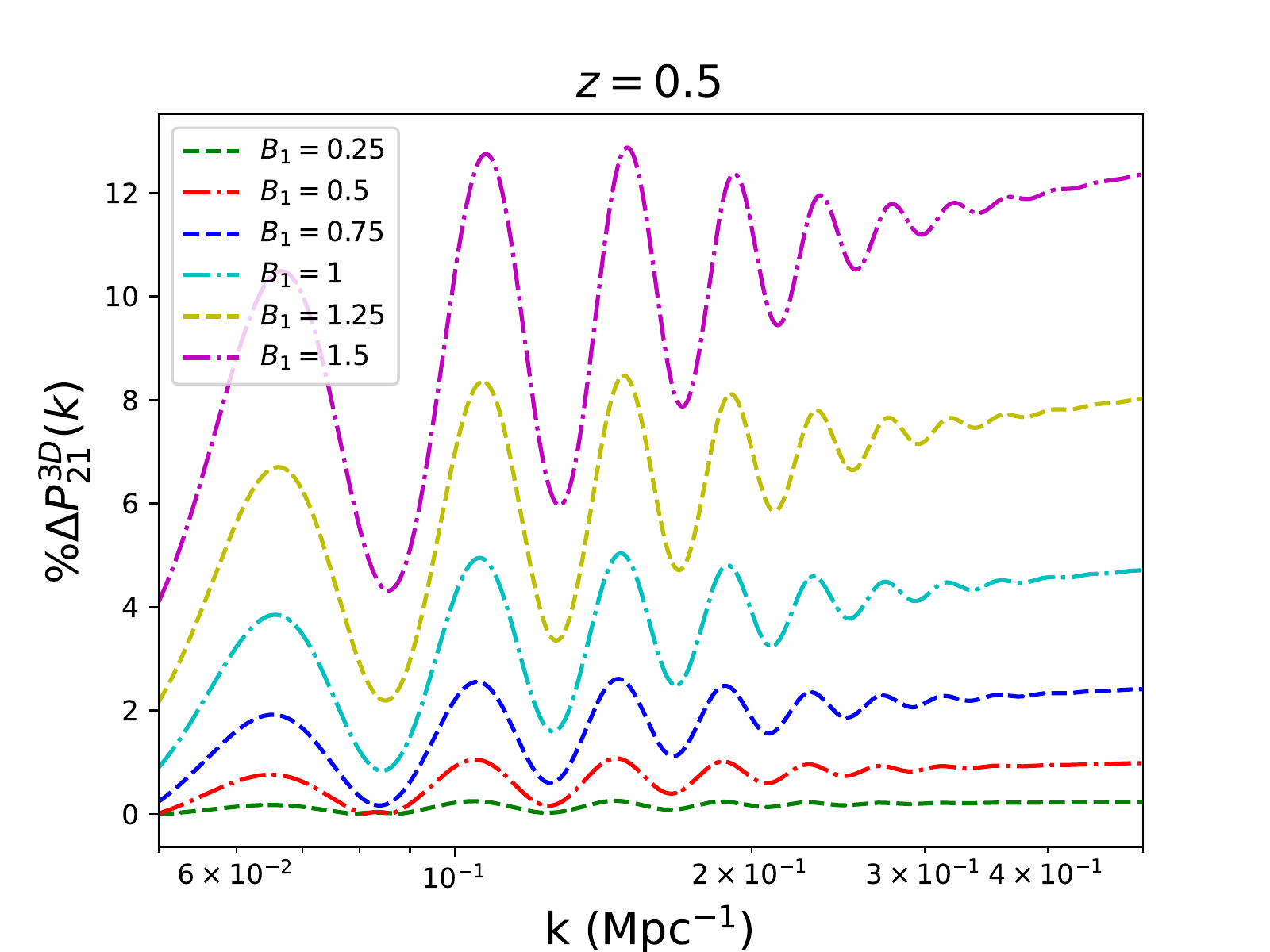}
  \includegraphics[width=0.45\textwidth]{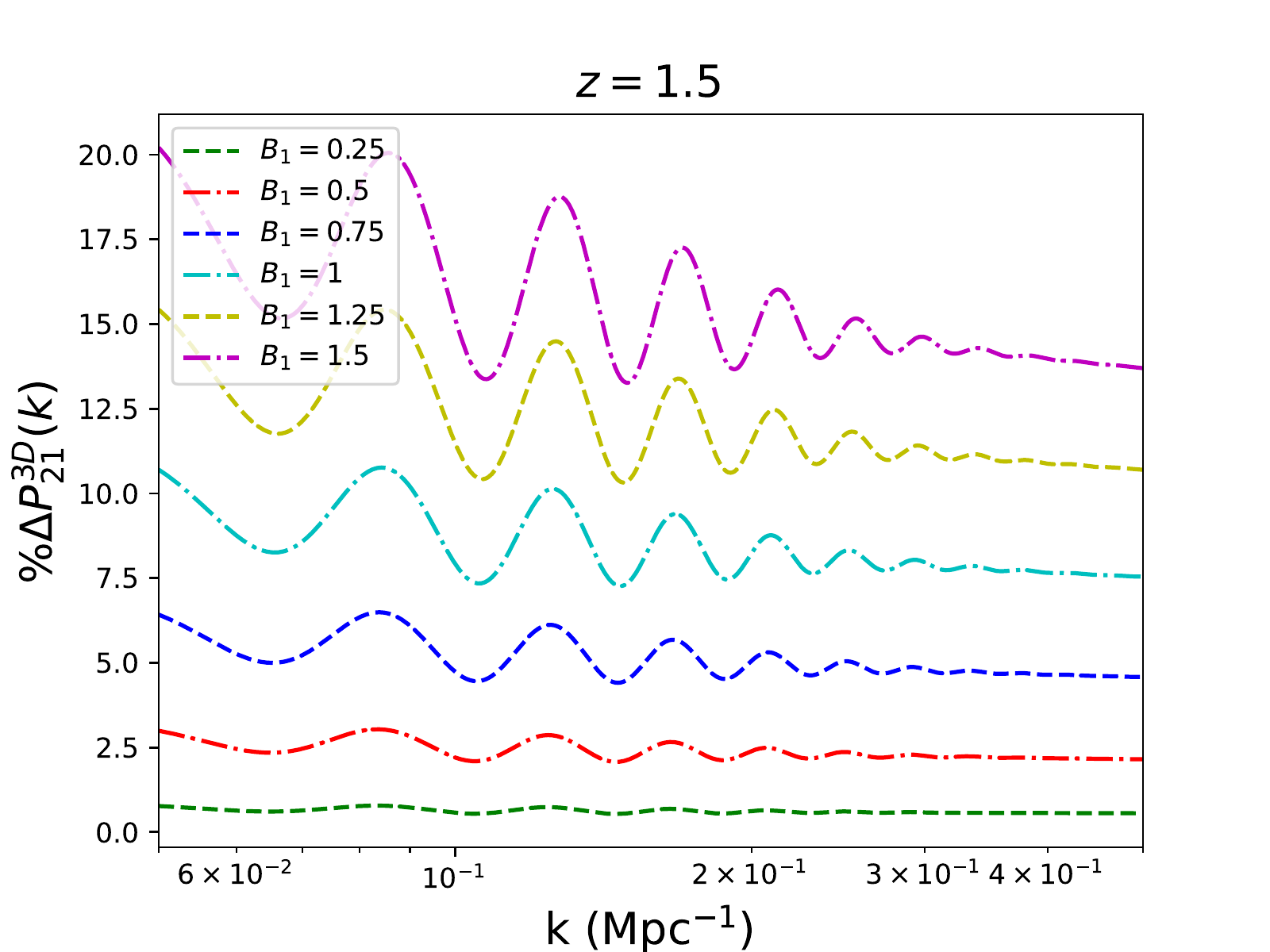}
  \includegraphics[width=0.45\textwidth]{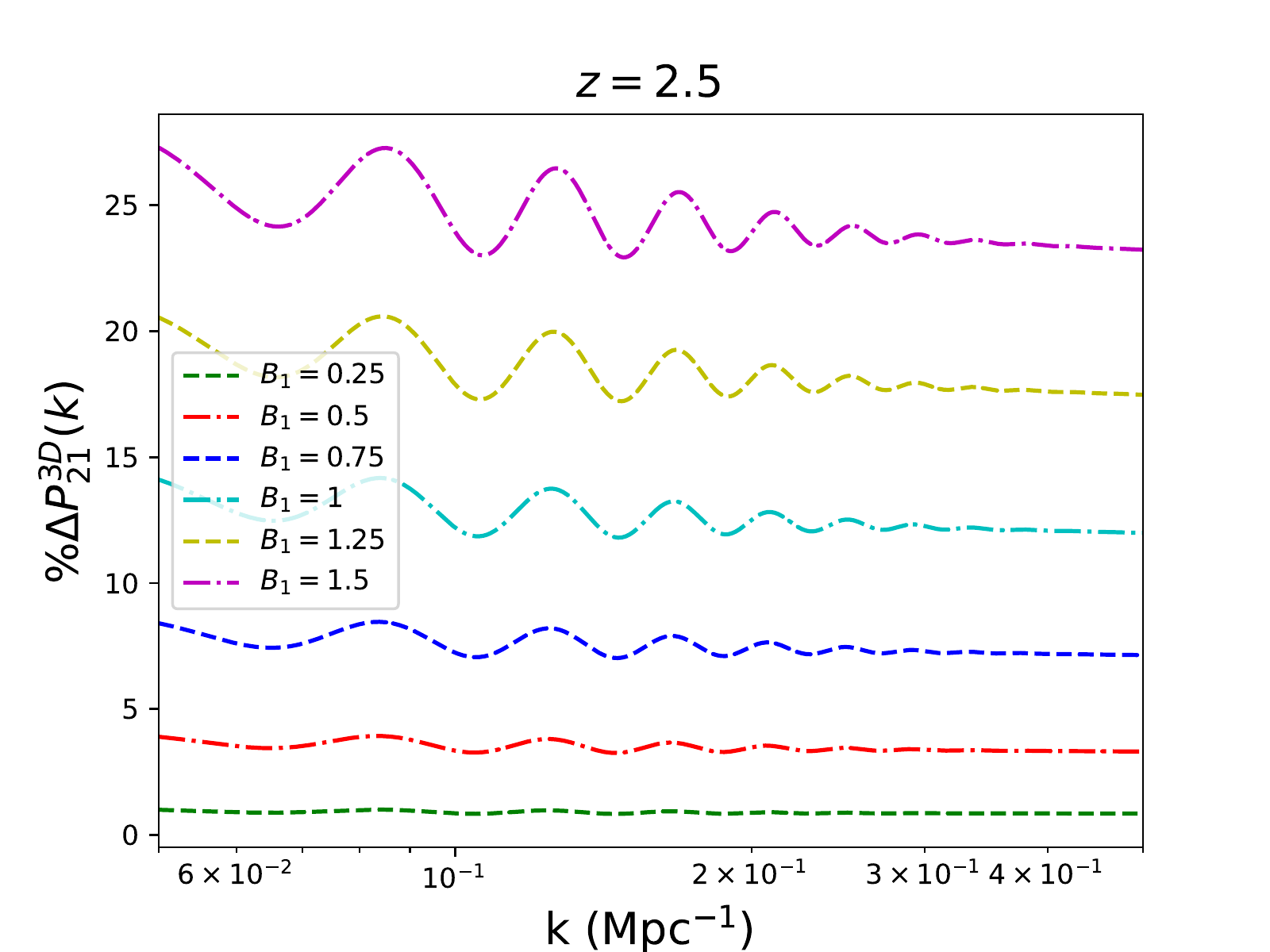}
  \caption{Percentage deviation in the 3D 21 cm power spectrum for the bimetric gravity model from the $\Lambda$CDM model. We have plotted  $\%\Delta P^{3D}_{21}$ w.r.t. $k$ at different redshifts for different values of parameter $B_1$.} 
  \label{fig:3d_21_plot}
\end{figure}

In Fig.~\ref{fig:3d_21_plot}, we show the absolute percentage deviation of the $\mu$ averaged 3D 21 cm power spectrum from the $\Lambda$CDM model. We see deviation up-to 25\%, depending upon parameter $B_1$, at different redshifts. We see that deviation in 3D 21 cm power spectrum is k-scale dependent because of the corrections arising due to the difference in the real cosmological model and the fiducial cosmological model.

\section{Prospect of Detectability of Bimetric Gravity with SKA1-Mid Telescope}
\label{sec-ska1mid}

We study the prospect of detectability of bimetric gravity using the upcoming SKA1-Mid telescope. The SKA1-Mid observation will have some observational errors. These errors are mainly of two types: one is the system noise, another one is the sample variance. Here, we neglect other errors like astrophysical residual foregrounds. So, we consider the detectability of bimetric gravity after the foreground removal from SKA1-Mid observation. For the expressions of system noise and sample variance, we closely follow \cite{Villaescusa-Navarro:2014cma}.

We start with the SKA1-Mid telescope antenna distributions from the document \url{https://astronomers.skatelescope.org/wp-content/uploads/2016/09/SKA-TEL-INSA-0000537-SKA1_Mid_Physical_Configuration_Coordinates_Rev_2-signed.pdf}. This document suggests that the SKA1-Mid observation will have 133 SKA antennas with the addition of 64 MEERKAT antennas. So, the total number of antennas, $N_t$ is 197. We assume this antenna distribution is circularly symmetric i.e. the distribution, $\rho_{\rm ant}$ depends only on the distance, $l$ from the center. With this assumption, we compute the 2D baseline distribution by a convolution integral equation given as \cite{Villaescusa-Navarro:2014cma}

\begin{equation}
\rho_{\rm 2D} (U,\nu_{\rm 21}) = B_{\nu}(\nu_{\rm 21}) \int_{0}^{\infty} 2 \pi l dl \hspace{0.1 cm} \rho_{\rm ant} (l)  \int_{0}^{2 \pi} d\phi \hspace{0.1 cm} \rho_{\rm ant} (|\vec{l}-\lambda_{\rm 21} \vec{U}|),
\label{eq:rho2d}
\end{equation}

\noindent
where $\nu_{\rm 21}$ and $\lambda_{\rm 21}$ are the observed frequency and wavelength of the 21 cm signal. $\vec{l}$ is the vector corresponding to distance, $l$. $\vec{U}$ is the baseline vector which is related to the wavevector, $\vec{k}$ given as $\vec{U} = \frac{ \vec{k}_{\perp} r }{2 \pi}$, where $\vec{k}_{\perp}$ is the wavevector in the transverse direction and $r$ is the comoving distance. $U$ is the magnitude of vector $\vec{U}$. $\phi$ is the angle between $\vec{l}$ and $\vec{U}$. $B_{\nu}$ is a normilization constant determined as

\begin{equation}
\int_{0}^{\infty} U dU \int_{0}^{\pi} d\phi \hspace{0.1 cm} \rho_{\rm 2D} (U,\nu_{\rm 21}) = 1.
\label{eq:rho2dnorm}
\end{equation}

\noindent
$B_{\nu}$ is a function of $\nu_{\rm 21}$ and consequently it is a function of $z$.

The 2D baseline distribution is not directly related to the system noise. Instead, the 3D baseline distribution is closely related to the system noise. The 3D baseline distribution, $\rho_{\rm 3D}$ is computed from the 2D baseline distribution given as

\begin{equation}
\rho_{\rm 3D}(k,\nu_{\rm 21}) = \left[ \int_{0}^{1} d\mu \hspace{0.2 cm} \rho_{\rm 2D}^{2} \left( \frac{r k}{2 \pi} \sqrt{1-\mu^{2}},\nu_{\rm 21} \right) \right] ^{\frac{1}{2}},
\label{eq:rho3d}
\end{equation}

\noindent
where $k$ is the magnitude of $\vec{k}$ and $\mu = \cos{\theta}$, where $\theta$ is the angle between $\vec{k}$ and the line of sight direction, as mentioned previously.

Another quantity that is related to the system noise is the total number of independent modes between $k$ to $k+dk$, and it is denoted as $N_k$. It is given as

\begin{equation}
N_k(k) = \frac{2 \pi k^2 dk}{V_{\rm 1-mode}},
\label{eq:Nk}
\end{equation}

\noindent
where $V_{1-mode}$ is the volume occupied by one independent mode given as

\begin{equation}
V_{\rm 1-mode} = \frac{(2 \pi)^{3} A}{r^2 L \lambda_{\rm 21}^{2}},
\label{eq:V_1_mode}
\end{equation}

\noindent
where $L$ is the comoving length corresponding to the bandwidth, $B$ of the observed signal, and $A$ is the physical collecting area of an antenna. We consider $A \approx 1256.6 m^2$ for a SKA1-Mid antenna.

Similarly, the number of independent modes lies between $k$ to $k+dk$ and $\theta$ to $\theta+d\theta$ is denoted by $N_m$ and it is given as

\begin{equation}
N_{m}(k,\theta) = \frac{2 \pi k^{2} dk \sin{\theta} d\theta}{V_{\rm 1-mode}}.
\label{eq:Nm}
\end{equation}

The system noise, $\delta P_{N}$ is given as \cite{Villaescusa-Navarro:2014cma,Sarkar:2015jta,Geil2010PolarizedFR,McQuinn:2005hk,Dinda:2018uwm,Hotinli:2021xln,Dinda:2022ixi}

\begin{equation}
\delta P_{N} (k,\nu_{\rm 21}) = \frac{T_{\rm sys}^2}{B t_0} \left( \frac{\lambda_{\rm 21}^2}{A_e} \right)^2 \frac{2 r^2 L}{N_t (N_t-1) \rho_{\rm 3D} (k,\nu_{\rm 21})} \frac{1}{\sqrt{N_k(k)}},
\label{eq:pnfinal}
\end{equation}

\noindent
where $T_{\rm sys}$ is the system temperature, $t_0$ is the observation time and $A_e$ is the effective collecting area of an antenna given as $A_e = \epsilon A$, where $\epsilon$ is the efficiency. We assume $\epsilon$ to be $0.7$. We consider the $T_{\rm sys}$ would be typically 40 Kelvin. We consider the observation time to be 1000 Hours.

The sample variance, $\delta P_{\rm SV}$ is given as \cite{Villaescusa-Navarro:2014cma,Sarkar:2015jta,Geil2010PolarizedFR,McQuinn:2005hk,Dinda:2018uwm,Hotinli:2021xln,Dinda:2022ixi}

\begin{equation}
\delta P_{\rm SV} (k,\nu_{\rm 21}) = \left[ \sum_{\theta} \frac{N_m(k,\theta)}{P_{\rm 21}^2(k,\theta)} \right]^{- \frac{1}{2}},
\label{eq:sv}
\end{equation}

\noindent
where $P_{\rm 21}$ is the 21 cm power spectrum corresponding to a fiducial cosmological model. Note that the system noise is independent of any fiducial model but sample variance depends on it. For the fiducial model, we consider the $\Lambda$CDM model with the parameter values according to the 2018 results of the Planck mission, as mentioned previously \cite{Planck:2018vyg}.

The total noise, $\delta P_{\rm tot}$ is given as

\begin{equation}
\delta P_{\rm tot} (k,\nu_{\rm 21}) = \sqrt{ \delta P_N^2 (k,\nu_{\rm 21}) + \delta P_{\rm SV}^2 (k,\nu_{\rm 21}) } .
\label{eq:totalN}
\end{equation}

\begin{figure}[H]
  \centering
  \includegraphics[width=0.45\textwidth]{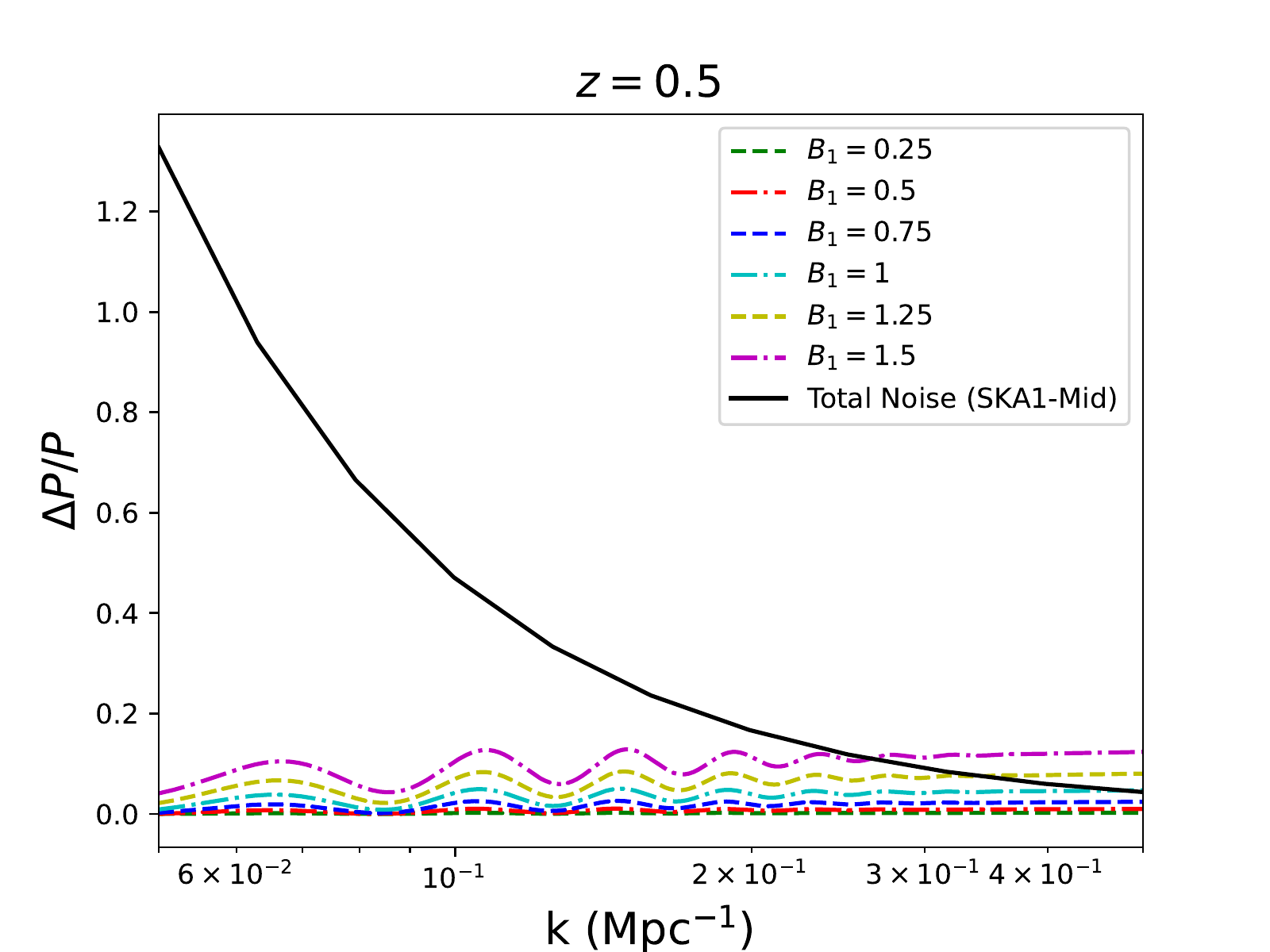}
  \includegraphics[width=0.45\textwidth]{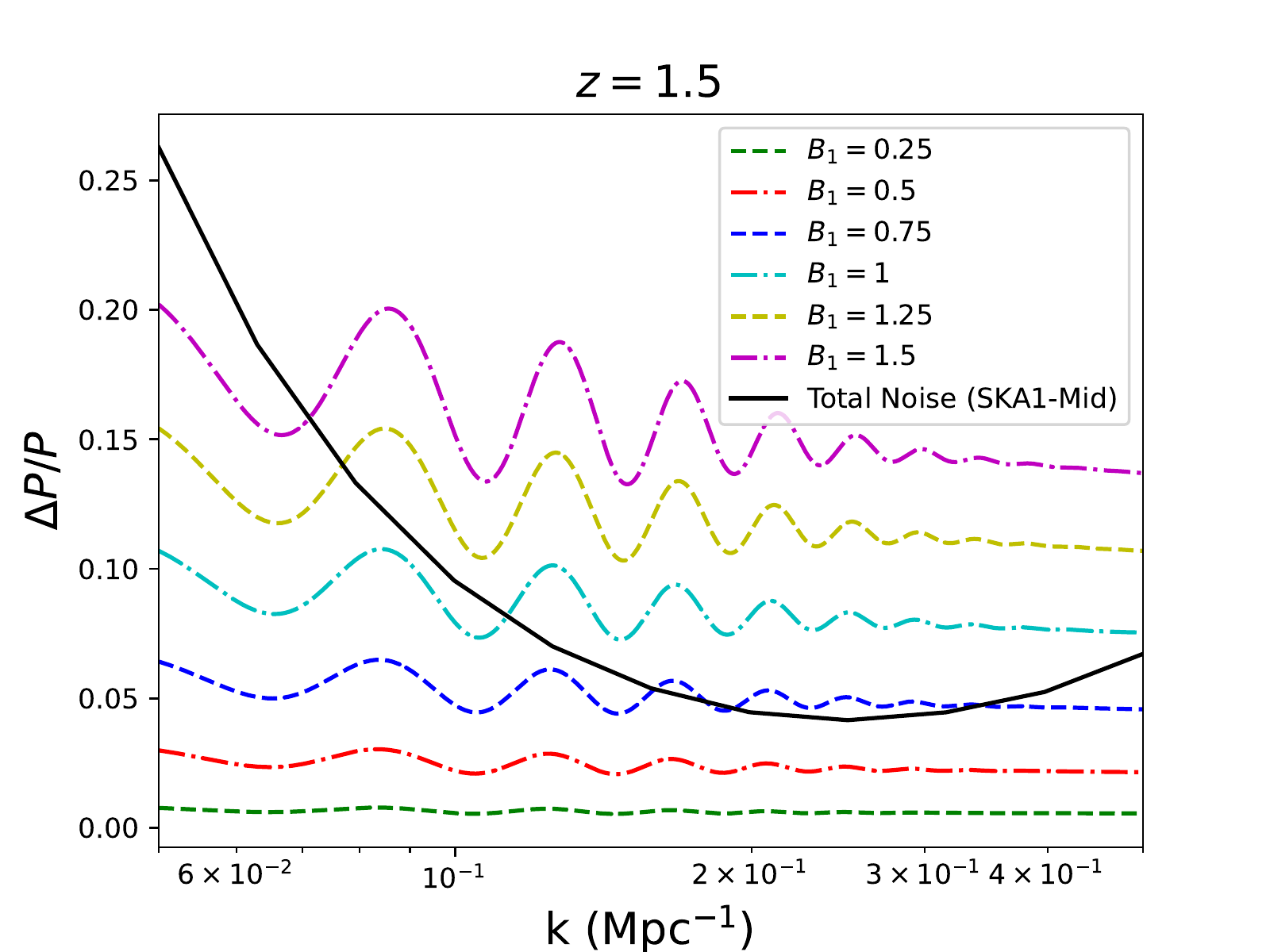}
  \includegraphics[width=0.45\textwidth]{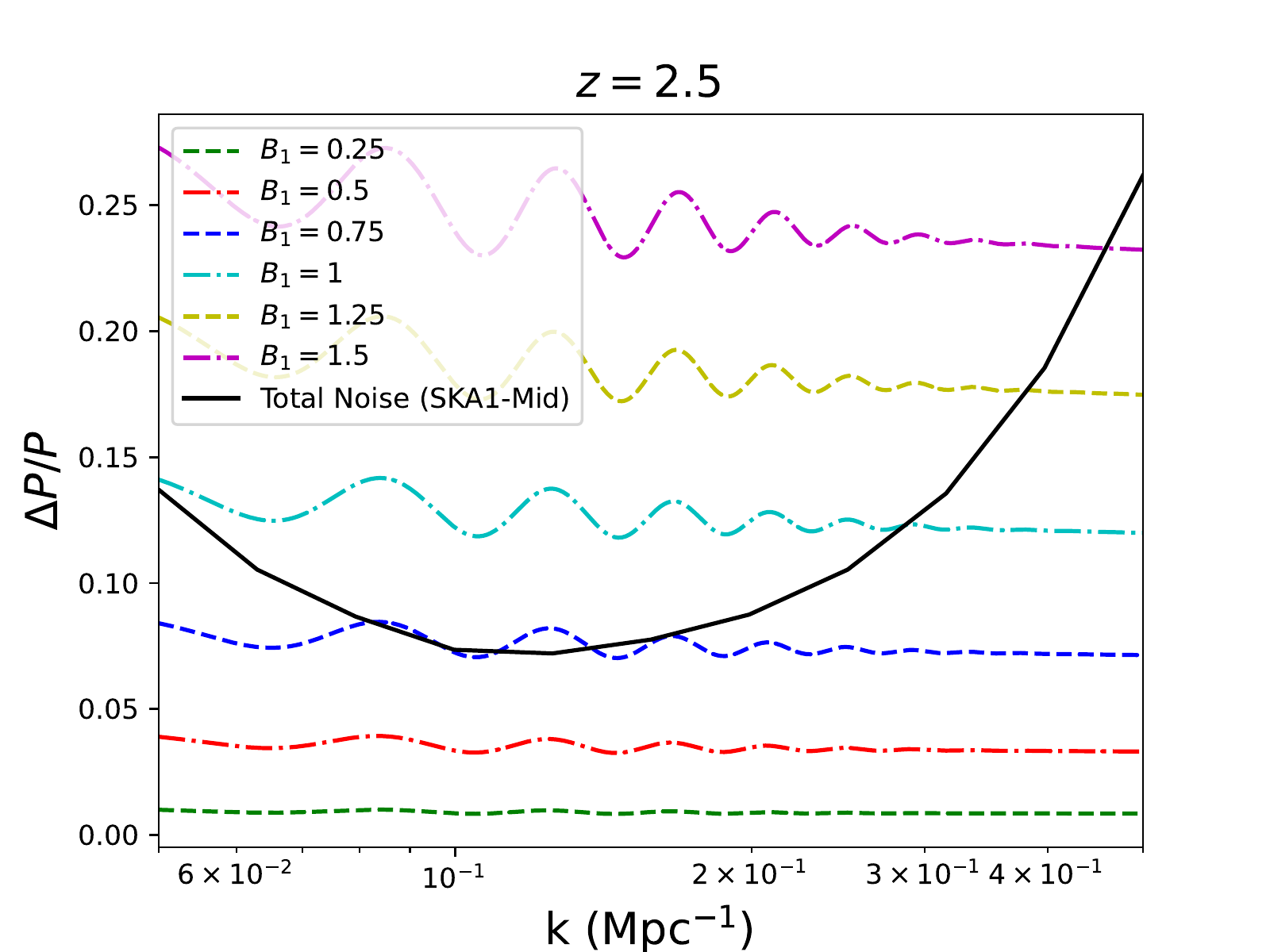}
\caption{
Detectability of the bimetric gravity model from the $\Lambda$CDM model using SKA1-Mid observation. Here, we have plotted $\delta P/P$ with $k$ at three different redshifts for different values of $B_1$. The top-left, top-right, and bottom panels correspond to redshifts $0.5$, $1.5$, and $2.5$ respectively. $P = P_{\rm fd} = P_{\rm 21}(\Lambda {\rm CDM})$. $\delta P = \left|P_{\rm 21}^{\rm 3D} - P_{\rm 21}(\Lambda {\rm CDM} )\right|$ for all the lines except black lines. For black lines $\delta P = \delta P_{\rm tot}$ obtained from Eq.~\ref{eq:totalN}.
} 
\label{fig:ska_21_plot}
\end{figure}
In Figure~\ref{fig:ska_21_plot}, we show how much we can detect the bimetric gravity with the upcoming SKA1-Mid telescope observation. In this figure, we have plotted $\delta P/P$ versus $k$, where $P = P_{\rm fd} = P_{\rm 21}(\Lambda {\rm CDM})$ is the 21 cm power spectrum corresponding to the fiducial model which is the $\Lambda$CDM here. $\delta P$ is the absolute deviation in the 3D 21 cm power spectrum corresponding to a particular theoretical model from the fiducial model. So, $\delta P = \left|P_{\rm 21}^{\rm 3D} - P_{\rm 21}(\Lambda {\rm CDM} )\right|$ (except for the black lines). Note that, for the observation i.e. for the black lines, the $\delta P = \delta P_{\rm tot}$ obtained from Eq.~\ref{eq:totalN}. Also, note that, for the fiducial model, the 3D 21 cm power spectrum is the same as the 2D 21 cm power spectrum. The models corresponding to the lines which are above these black lines are detectable by the upcoming SKA1-Mid observation.

\section{Conclusions}
\label{concl}
In the present work, we have used a subclass of ghost-free bimetric gravity as a modified theory of gravity to study the large scale structures by using the 21 cm power spectrum. We have used Raychaudhury's equation to study the growth of large scale structures up to linear order of perturbations for the bimetric gravity. We have studied the 21 cm power spectrum and its deviation from $\Lambda$CDM model for different values of parameter $B_1$. We observe that the 21 cm power spectrum is scale independent but as we consider the fiducial model of cosmology for the baseline distributions of interferometers, like SKA, the corrected 21 cm power spectrum developed a k-scale dependency due to difference between fiducial model and the bimetric gravity model. We have considered the SKA1-Mid interferometric observations to put constraints on the detectability of 21 cm power spectrum for bimetric gravity from the $\Lambda$CDM model. We have considered the system noise and the sample variance according to the SKA1-Mid observations with $\Lambda$CDM as the fiducial model. For an ideal observations, we haven't considered any other errors in the 21 cm power spectrum observations. We observe that the forthcoming SKA1-Mid observations can put strong constraints on the bimetric gravity, and on higher redshifts, we can have greater possibility to distinguish the bimetric gravity from the $\Lambda$CDM model.

\section*{Acknowledgments}
BRD would like to acknowledge IISER Kolkata for the financial support through the postdoctoral fellowship.  AAS acknowledges the funding from SERB, Govt of India under the research grant MTR/20l9/000599. 

\section*{References}

\bibliography{ref}

\bibliographystyle{ieeetr}

\end{document}